\documentclass[
reprint, 
superscriptaddress, 
amsmath, 
amssymb, 
aps, 
prb
]{revtex4-1}

\usepackage{graphicx}
\usepackage{dcolumn}
\usepackage{bm}

\begin{document}
\preprint{APS/123-QED}

\title{Hole-doping-induced melting of spin-state ordering in PrBaCo$_2$O$_{5.5+x}$}


\author{Ping Miao}
\email{miao@post.kek.jp}
\affiliation{Institute of Materials Structure Science, High Energy Accelerator Research Organization (KEK), Tokai 319-1106, Japan}
\affiliation{Sokendai (The Graduate University for Advanced Studies), Tokai 319-1106, Japan}

\author{Xiaohuan Lin}
\affiliation{College of Chemistry and Molecular Engineering, Peking University, Beijing 100871, China}

\author{Sanghyun Lee}
\affiliation{Institute of Materials Structure Science, High Energy Accelerator Research Organization (KEK), Tokai 319-1106, Japan}

\author{Yoshihisa Ishikawa}
\affiliation{Institute of Materials Structure Science, High Energy Accelerator Research Organization (KEK), Tokai 319-1106, Japan}

\author{Shuki Torii}
\affiliation{Institute of Materials Structure Science, High Energy Accelerator Research Organization (KEK), Tokai 319-1106, Japan}

\author{Masao Yonemura}
\affiliation{Institute of Materials Structure Science, High Energy Accelerator Research Organization (KEK), Tokai 319-1106, Japan}
\affiliation{Sokendai (The Graduate University for Advanced Studies), Tokai 319-1106, Japan}

\author{Tetsuro Ueno}
\affiliation{Institute of Materials Structure Science, High Energy Accelerator Research Organization (KEK), Tokai 319-1106, Japan}
\affiliation{National Institute for Materials Science, Tsukuba, Ibaraki 305-0047, Japan}

\author{Nobuhito Inami}
\affiliation{Institute of Materials Structure Science, High Energy Accelerator Research Organization (KEK), Tokai 319-1106, Japan}

\author{Kanta Ono}
\affiliation{Institute of Materials Structure Science, High Energy Accelerator Research Organization (KEK), Tokai 319-1106, Japan}
\affiliation{Sokendai (The Graduate University for Advanced Studies), Tokai 319-1106, Japan}

\author{Yingxia Wang}
\affiliation{College of Chemistry and Molecular Engineering, Peking University, Beijing 100871, China}

\author{Takashi Kamiyama}
\email{takashi.kamiyama@kek.jp}
\affiliation{Institute of Materials Structure Science, High Energy Accelerator Research Organization (KEK), Tokai 319-1106, Japan}
\affiliation{Sokendai (The Graduate University for Advanced Studies), Tokai 319-1106, Japan}


\date{\today}

\begin{abstract}
The layered perovskite cobaltite RBaCo$_2$O$_{5.5}$ (R: rare-earth elements or Yttrium) exhibits an abrupt temperature-induced metal$-$insulator transition (MIT) and has been attributed to spin-state ordering (SSO) of Co$^{3+}$ ions. Here we investigated the hole doping member of PrBaCo$_2$O$_{5.5+x}$ ($0 \le x \le 0.24$) with multiple techniques. The analysis on crystal and magnetic structures by electron and neutron diffraction confirm the SSO in the insulating phase of undoped PrBaCo$_2$O$_{5.5}$, which is melted by increasing the temperature across the MIT. In addition, we discovered that hole doping to PrBaCo$_2$O$_{5.5}$ also melts the SSO in conjunction with an insulator-metal transition. The experimental results from electron/neutron diffraction and soft x-ray absorption spectroscopy (XAS) all lead to the conclusion that hole-doping induced MIT occurs is in the same manner as the temperature-induced MIT. Therefore, we propose a unified mechanism that dominates the temperature- and hole-doping-induced MITs in the PrBaCo$_2$O$_{5.5+x}$ system. Specifically, this mechanism involves symmetry breaking coupled with a SSO in the paramagnetic phase.
\end{abstract}


\maketitle

\section{Introduction}
Metal$-$insulator transition  transition (MIT) is observed in many transition-metal oxides and manifests itself by large change in resistivity across a phase boundary. Interplay between the spin/charge/orbital/lattice degrees of freedom has been found to play a crucial role in developing various electronic phases that span both the metallic and insulating sides, e.g., ferromagnetic metal, antiferromagnetic metal, Mott$-$Hubbard insulator, charge-transfer insulator.\cite{MIT_1968, MIT_1998} Recently, another type of degree of freedom has drawn much attention for its effects on the magnetic and transport properties of iron-based superconductors.\cite{Iron_spin_state_2013, Iron_spin_state_prl_2013, Iron_spin_state_2011} It is referred to as the spin state degree of freedom (SSDF) and describes the distribution of d-electronic spins over the crystal-field-split $t_{2g}$ and $e_g$ orbitals. 
\begin{figure}
\includegraphics[width=0.8\columnwidth]{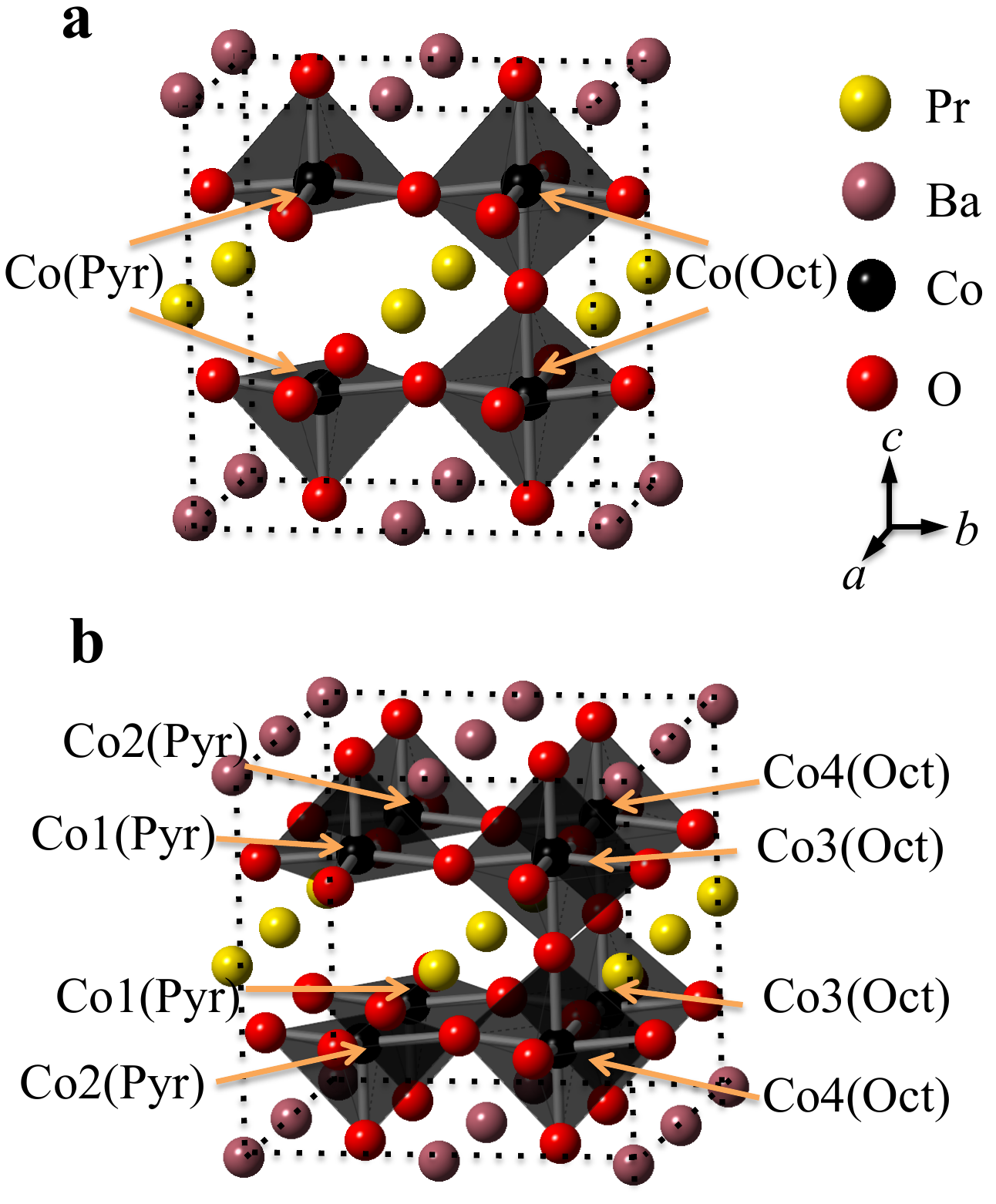}
\caption{\label{fig1}  (a)  The  $Pmmm (a_p \times 2a_p \times 2a_p)$ cyrstal unit cell with oxygen vacancy in the PrO$_x$ layer leading to ordering along the $b$ axis. The crystal structure consists of two nonequivalent crystallographic Co sites: one pyramidal site Co(Pyr) and one octahedral site Co(Oct). (b)  $Pmma (2a_p \times 2a_p \times 2a_p)$ cyrstal unit cell with four nonequivalent crystallographic Co sites, labeled Co1(Pyr), Co2(Pyr), Co3(Oct) and Co4(Oct).}
\end{figure}

Although SSDF is present in many $d$-electron systems, how it correlates with the properties of materials is still not well understood. LaCoO$_3$ has been known to be a prototypic system for exhibiting different spin states of the trivalent cobalt atoms through three kinds of configuration of 3$d^6$ electrons: low-spin (LS, $t_{2g}^6e_g^0$, $S$ = 0), intermediate-spin (IS, $t_{2g}^5e_g^1$, $S$ = 1) and high-spin (HS, $t_{2g}^4e_g^2$, $S$ = 2) states. Upon increasing temperature to around 100 K, it undergoes a spin-state transition (SST) in which the nonmagnetic LS ground-state transforms to a paramagnetic (PM) higher spin state, although whether this PM susceptibility stems from HS or IS state is still debated over.\cite{MIT_1998, LaCoO3_1953, LaCoO3_prl_2013, LaCoO3_prl_2009}  Goodenough $et$ $al.$\cite{LaCoO3_1958, LaCoO3_1967} postulated that a long-range ordering of LS and HS Co ions, which is called spin-state ordering (SSO) (not to be confused with the ordering of magnetic moments in traditional magnetically ordered structures), accompanies the SST and elucidates the magnetic and charge transport properties. To be noted, SST is defined by a change of the spin states, so it is not necessarily to be accompanied by SSO, whereas SSO always occurs together with a SST. However, static ordering of LS and HS (or IS) ions in bulk LaCoO$_3$ has never been found experimentally.\cite{LaCoO3_1967, LaCoO3_1999, LaCoO3_2013, LaCoO3_1986, LaCoO3_2000} On the other hand, Korotin $et$ $al.$\cite{LaCoO3_1996} demonstrated by LDA+U calculations that the IS state is almost degenerate with the LS ground state and lower in energy than the HS state. This IS scenario has been popular until recently several theoretical studies\cite{LaCoO3_prl_2013, LaCoO3_2012, LaCoO3_prb_2009, LaCoO3_2011, LaCoO3_zhang_2012} revived the HS scenario in terms of the dynamically or statically mixed LS/HS state, which is corroborated by the result from x-ray absorption spectroscopy.\cite{LaCoO3_2006}  However, the latest inelastic x-ray scattering (IXS) and infrared spectroscopy experiments\cite{LaCoO3_2014} indicates a complex spin-state disproportionation including all the LS, IS, and HS states.

Much attention has been paid to an analogous cobaltite family, RBaCo$_2$O$_{5.5}$ (R: Nd\cite{Fauth_2002}, Gd\cite{Chernenkov_2005}, Tb\cite{Plakhty_2005}, Dy\cite{Chernenkov_2007}, Ho\cite{Jorgensen_2008} and Y\cite{Khalyavin_2007}) with the nominal valence of Co ion being +3 because a long-range SSO has been found to occur in conjunction with a temperature-induced (TI) MIT. The poor metallic phase of RBaCo$_2$O$_{5.5}$ crystalizes into an orthorhombic crystal structure (space group $Pmmm$) with a unit cell $a_p \times 2a_p \times 2a_p$, where $a_p$ is the lattice parameter of the pseudocubic perovskite ABO$_3$ subcell,\cite{Maignan_1999, Kusuya_2001, Frontera_2002} see Fig.~ \ref{fig1}(a). The ordered distribution of oxygen vacancies in the PrO$_x$ layer results in an alternative arrangement of corner-sharing CoO$_5$ pyramids (Pyr) and CoO$_6$ octahedra (Oct) along the $b$ axis, leading to two nonequivalent crystallographic sites, labeled Co(Pyr) and Co(Oct). Some of early studies\cite{Kusuya_2001, Frontera_2002}  asserted that the $Pmmm (a_p \times 2a_p \times 2a_p)$ structure survives when the material undergoes a TI-MIT into the insulating state. Later, Chernenkov $et$ $al.$\cite{Chernenkov_2005} used single-crystal x-ray diffraction to reveal superlattice reflections along the $a$ axis below $T_{MI}$, which are absent in the poor-metallic phase in GdBaCo$_2$O$_{5.5}$, and thus concluded that the insulating phase consists of a different orthorhombic structure  $Pmma (2a_p \times 2a_p \times 2a_p)$ with four Co sites: Co1(Pyr), Co2(Pyr), Co3(Oct) and Co4(Oct) [see Fig.~\ref{fig1}(b)]. They attributed the splitting of two sites of $Pmmm$ structure into four sites of $Pmma$ structure the SSO. Following up works corroborated this model and assigned the specific spin states of these four sites through assessing the individual magnetic moment of the Co$^{3+}$ ions with the magnetic structures analyzed from neutron powder diffraction.\cite{Plakhty_2005, Jorgensen_2008, Khalyavin_2007, Chernyshov_2008, Khalyavin_2005}

A similar drop in resistivity that occurs as a function of the hole-doping fraction $x$ was found in the high temperature region of resistivity curve of single-crystal GdBaCo$_2$O$_{5.5+x}$\cite{Taskin_2005} and thin films of LnBaCo$_2$O$_{5.5+x}$ (Ln = Er, Pr)\cite{Bao_2014}, which we hereafter refer to as the hole-doping induced metal$-$nsulator transition (HDI-MIT). However, no reports have ever discuss the hole doping effect on the spin states involving the Co$^{3+}$ ions or the relationship between the HDI-MIT and the TI-MIT of the RBaCo$_2$O$_{5.5+x}$ system, partly because of lack of consensus on the detailed crystal and magnetic structures of RBaCo$_2$O$_{5.5+x}$ ($x \ge 0$).\cite{Chernenkov_2005, Plakhty_2005, Chernenkov_2007, Jorgensen_2008, Khalyavin_2007, Maignan_1999, Kusuya_2001, Frontera_2002, Fauth_2001, Soda_2003} To resolve this situation, we took advantage of the fact that PrBaCo$_2$O$_{5.5+x}$ can reach the highest hole-doping level ($0 \le x \le 0.24$) within the rare-earth family and carried out a multi-probe study (neutron, electron and x-rays) on the crystal and Co-ion spin/orbital structures as well as characterize the transport and magnetic properties of a full series of polycrystalline samples.

\begin{figure}
\includegraphics[width=0.8\columnwidth]{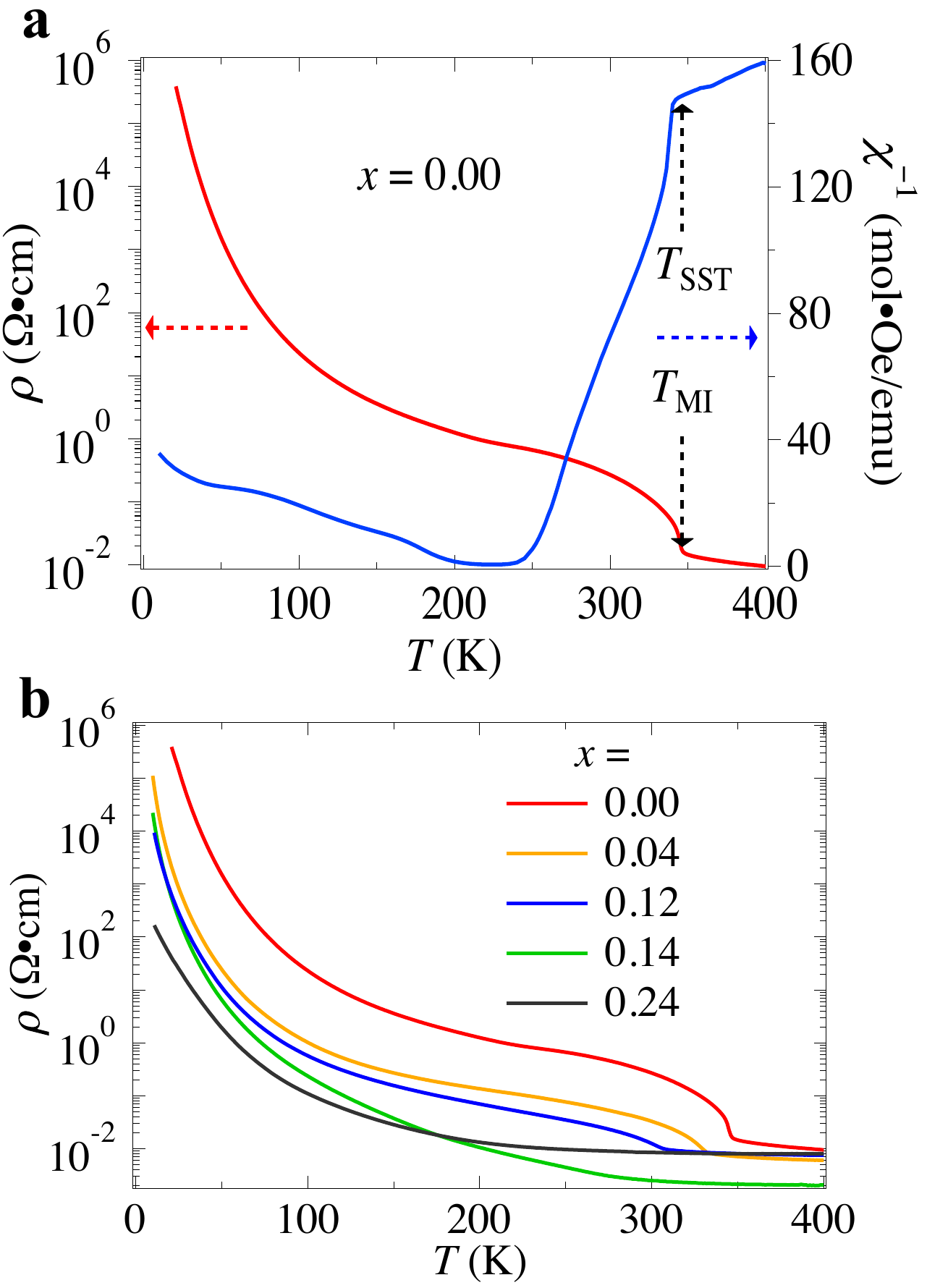}
\caption{\label{fig1_1} (a)  Resistivity $\rho$ and inverse magnetic susceptibility $\chi^{-1}$ as a function of temperature for sample with $x$ = 0. The abrupt changes in $\rho$ and $\chi^{-1}$ at $T_{MI}$ mark the concurrence of the SST and MIT. The susceptibility was measured in the 100 Oe field after zero-field cooling (ZFC). (b)  Resistivity $\rho$ as a function of temperature for various hole-doping fraction x, which shows a sharp drop with increasing hole doping.  The $x$-dependence of $\rho$ reveals an HDI-MIT at $x$ = 0.12. The anomalously low value in $\rho$ for $x$ = 0.14 above $\sim$200 K is possibly due to the lower boundary resistance of polycrystaline grains compared with that in samples with other hole-doping fractions. This speculation can be verified by the resistivity as a function of $x$ in single crystal of GdBaCo$_2$O$_{5.5+x}$.\cite{Taskin_2005}}
\end{figure}
\section{Experimental Method}
PrBaCo$_2$O$_{5.5+x}$ polycrystalline samples were synthesized by the solid-state reaction method with a combined EDTA$-$citrate complex sol$-$gel process.\cite{Synthesis_2008, Synthesis_2007} The hole-doping fraction $x$ was controlled by annealing the as-prepared samples in pure nitrogen atmosphere at various temperatures with the nominal $x$ ranging from 0 to 0.24.  The fraction $x$ was determined by both iodometric titration and neutron powder diffraction, with the results agreeing to each other within the error of $\sim$0.02. The high-resolution neutron powder diffraction measurements were performed using SuperHRPD at Japan Proton Accelerator Research Complex (J-PARC). The best resolution is $\frac{\Delta d}{d}$ = 0.035\% at 2$\theta$ of 172$^{\circ}$. We used the software $SARAh$\cite{Wills_2000} to carry out the representation analysis on magnetic structure. Rietveld refinement was performed using the software $Z$-$Rietveld$\cite{Rietveld_2009, Rietveld_2012} and $FullProf$\cite{Rietveld_1993},  and the residual values R$_{wp}$ and R$_{M}$ are both below 10\%. The electron diffraction pattern was acquired on a JEOL JEM-2100 transmission electron microscope with an accelerating voltage of 200 kV and a camera length of 150 cm, giving a point resolution of 2.3 \AA. Soft x-ray absorption spectroscopy (XAS) was carried out at the BL-16A of the Photon Factory in High Energy Accelerator Research Organization (KEK). Samples were mounted in a liquid-He cryostat and XAS spectra were obtained in total electron yield (TEY) mode.  The Resistivity was measured using a standard dc four-probe method on a Quantum Design Physical Property Measurement System (PPMS), and the magnetization was measured using a dc superconducting quantum interference device (SQUID) magnetometer (MPMS) at the Cross-Tokai user laboratories.
\section{Results}

\subsection{TI-MIT and HDI-MIT}

In the undoped member, PrBaCo$_2$O$_{5.5}$, the TI-MIT occurring in paramagnetic (PM) phase can be identified by a sharp drop in resistivity $\rho$ around $T_{MI}$  = 350 K, which corresponds to the transition from an insulator to a poor metal [see  Fig.~\ref{fig1_1}(a)]. The change of slope of the inverse magnetic susceptibility in the PM phase in Fig.~ \ref{fig1_1}(a) corresponds to the Curie-Weiss effective magnetic moments $\mu_{eff}$ = 2.50(1) $\mu_B$ and 5.39(1) $\mu_B$ below and above $T_{MI}$, respectively. Therefore, the SST involving the Co$^{3+}$ ions is evident from such a significant difference on assumption of the same contribution from Pr$^{3+}$ ions to the effective magnetic moments. HDI-MIT is a similar drop more than two order of magnitude in resistivity that occurs as a function of hole-doping fraction $x$ [see $\rho$ vs $x$ of Fig. \ref{fig1_1}(b)].

\subsection{Crystal and magnetic structures of PrBaCo$_2$O$_{5.5}$}
\begin{figure}
\includegraphics[width=0.8\columnwidth]{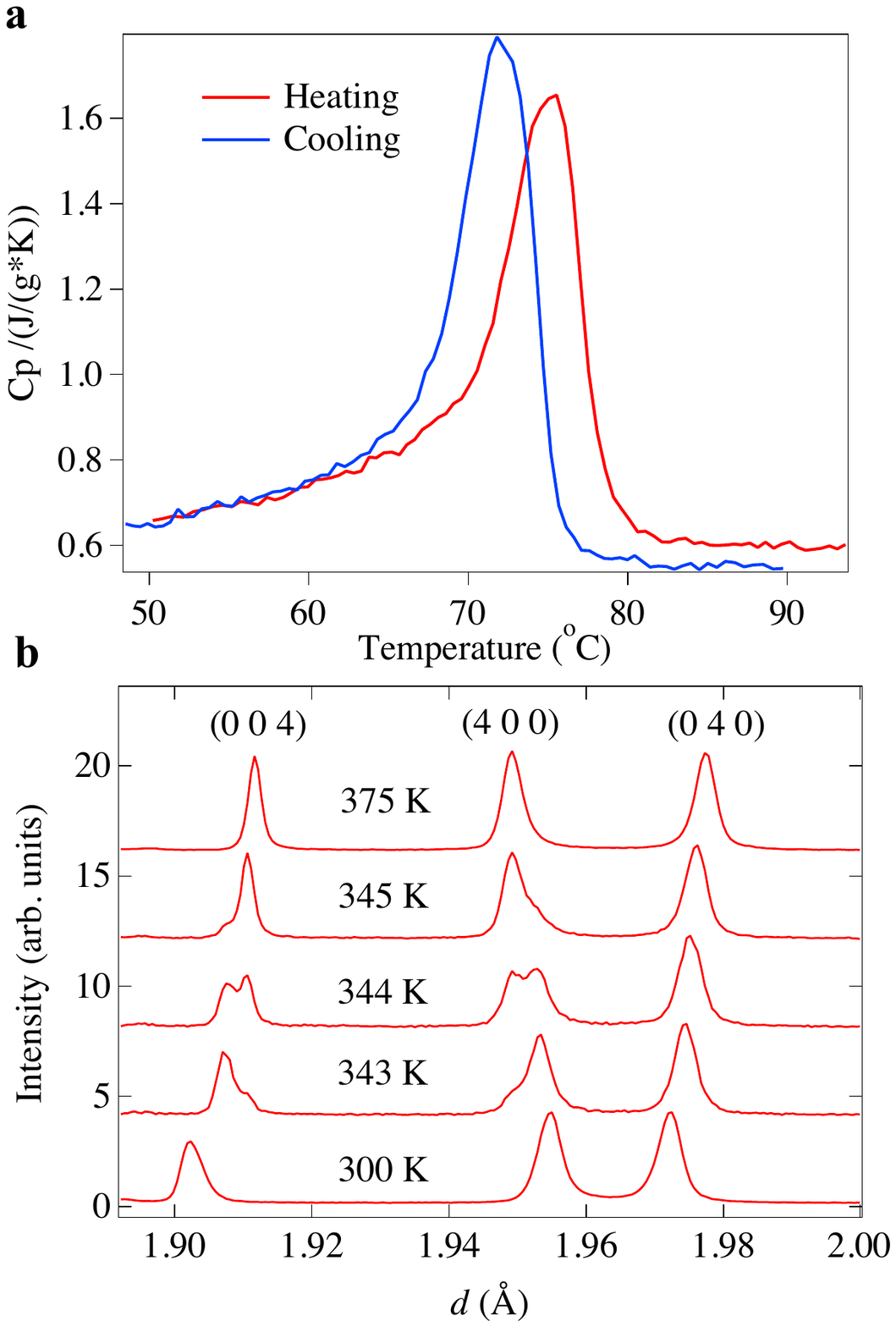}
\caption{\label{fig2}  (a) Specific heat curve in both heating and cooling process. The sharp peak and hysteresis between heating and cooling process indicate a first order phase transition. (b)  High-resolution neutron powder diffraction pattern, where the bragg peaks are indexed by the $(2a_p \times 2a_p \times 2a_p)$ unit cell. The coexistence of two phases in the critical region is a clear indication of a first order phase transition.}
\end{figure}
%

\begin{figure*}
\includegraphics[width=0.8\textwidth]{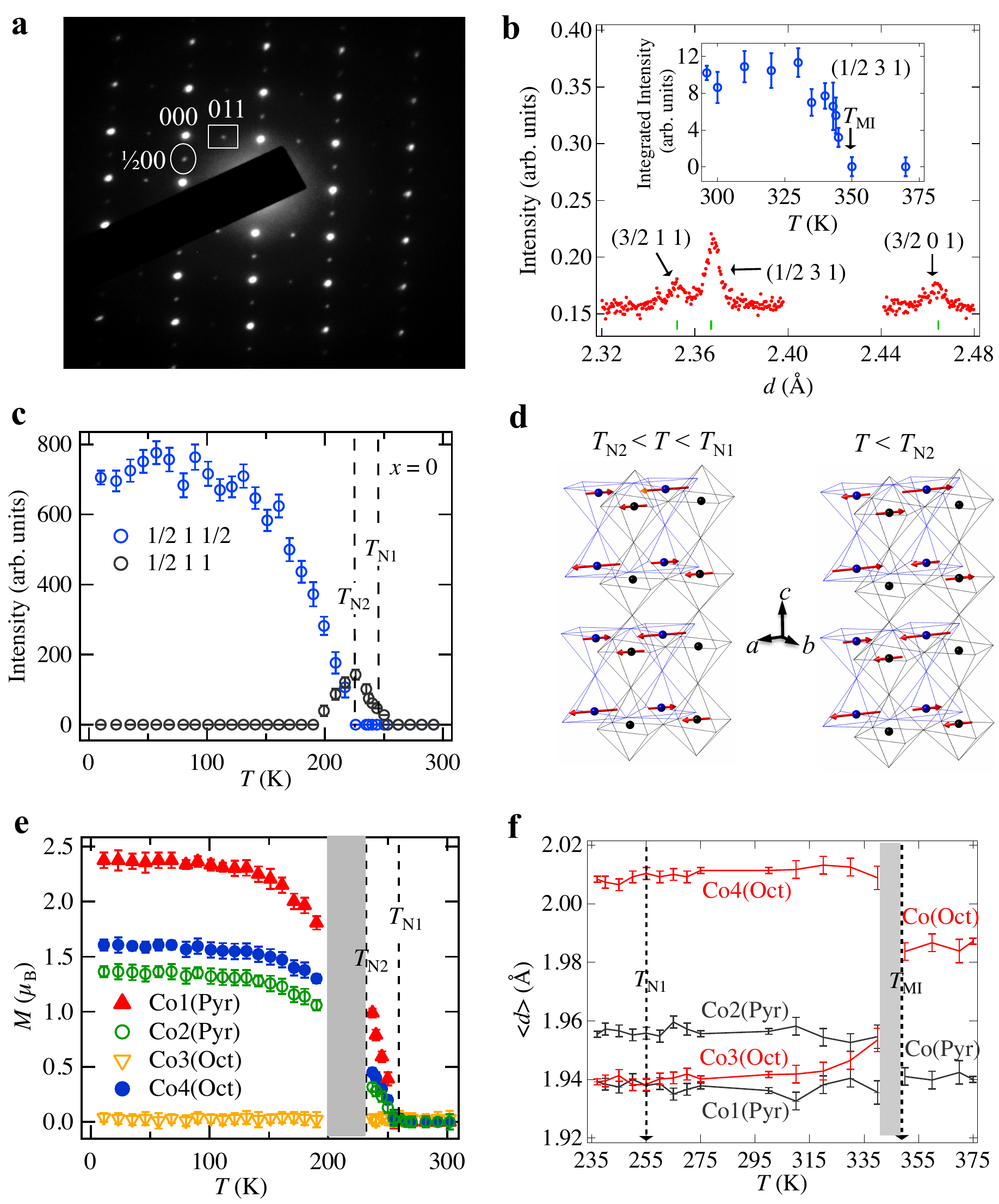}
\caption{\label{fig3}  Results of electron diffraction (a) and pulsed neutron powder diffraction (b, c, d, e) for PrBaCo$_2$O$_{5.5}$ (TI-MIT). (a)  Electron diffraction pattern at 300 K, which shows a superlattice reflection along the a axis with respect to  $(a_p \times 2a_p \times 2a_p)$ unit cell; $i.e.$, the 011 reflection is the spot in rectangle and the $\frac{1}{2}00$ reflection is the spot in circle. We verified that no multiple scattering occurred during the measurement. (b) The superlattice reflections  $\frac{3}{2}11$, $\frac{1}{2}31$, and $\frac{3}{2}01$  indexed  by the $(a_p \times 2a_p \times 2a_p)$  unit cell observed by neutron powder diffraction at 300 K (below  $T_{MI}$). (c)  Integrated intensity as a function of temperature for magnetic reflections magnetic reflections 111 and $11\frac{1}{2}$ indexed by the unit cell $(2a_p \times 2a_p \times 2a_p)$. (d) Ferrimagnetic ($T_{N2} \le T \le T_{N1}$) structure of Phase 1 and antiferromagnetic ($T \le T_{N2}$) structures of Phase 2. Octahedra are in black and pyramids in blue.(e)  Magnitude of magnetic moment as a function of temperature for each Co site in Phase 1, as determined from Rietveld refinement with the magnetic model in (d). The shaded region corresponds to the coexisting ferrimagnetic and antiferromagnetic structures. (f) Average Co$-$O bond length $<d>$ as a function of temperature for each Co site in the insulating phase (below $T_{MI}$) of $Pmma$ structure and in the poor metallic phase (above $T_{MI}$) of $Pmmm$ structure. The shaded region corresponds to the coexisting structures [see Fig. \ref{fig2}(b)]. The red (black) curves stands for CoO$_6$ octahedron (CoO$_5$ pyramid).}
\end{figure*}

Because the SSO model has not been reported for PrBaCo$_2$O$_{5.5}$, we first the study the crystal structure and the related magnetic structure to verify the SSO in this compound.  As shown in Fig.~\ref{fig2}, the TI-MIT in PrBaCo$_2$O$_{5.5}$ is identified to be first-order phase transition by both the specific heat measurement and high-resolution neutron powder diffraction (NPD).  The high-resolution NPD also allows us to obtain a sufficiently high signal-to-noise ratio to detect the superlattice reflections ($\frac{3}{2} 1 1$, $\frac{1}{2} 31$, and $\frac{3}{2} 01$) in the insulating phase [see Fig.~\ref{fig3}(b)]. Furthermore, electron diffraction, a sensitive technique in detecting weak superlattice reflections also provides corroborative evidence, as shown in Fig.~\ref{fig3}(a). These results confirm that the PrBaCo$_2$O$_{5.5}$ crystal structure undergoes a first-order phase transition  from high-temperature $Pmmm (a_p \times 2a_p \times 2a_p)$ to low-temperature $Pmma (2a_p \times 2a_p \times 2a_p)$, which is coincident with the TI-MIT. The results from Rietveld refinement for the $Pmma$ and $Pmmm$ structures are shown in Table~\ref {coordinate550pmma} and Table~\ref {coordinate550pmmm}, respectively.

\begin{table*}
\caption{\label{coordinate550pmma}  The atomic occupancy, coordinates and the isotropic displacement parameters for $x$ = 0 sample obtained from Rietveld refinement on neutron powder diffraction data at 300 K. The refinement is performed using orthorhombic $Pmma (2a_p \times 2a_p \times 2a_p)$ structure.The lattice parameters and criteria of the refinement quality are: $a$ = 7.81837(3) \AA, $b$ = 7.88788(3) \AA, $c$ = 7.60990(2) \AA; $R_{wp}$ =  8.70\%. }
\begin{ruledtabular}
\begin{tabular}{ccccccc}
Atom&Site&Occupancy&$x$&$y$&$z$&$100U_{iso}(\AA^2)$\\ \hline
Ba&$4g$&1&0&0.2499(1)&0&0.28(1)\\
Pr&$4h$&1&0&0.2706(1)&0.5&0.69(1)\\
Co1&$2e$&1&0.25&0&0.2502(3)&0.66(4)\\
Co2&$2e$&1&0.75&0&0.2511(3)&0.33(4)\\
Co3&$2f$&1&0.25&0.5&0.2516(3)&0.15(4)\\
Co4&$2f$&1&0.75&0.5&0.2495(3)&0.89(5)\\
O1&$2e$&1&0.25&0&0.9994(2)&1.37(1)\\
O2&$2f$&1&0.25&0.5&0.9970(2)&0.38(1)\\
O3&$4i$&1&0.0111(1)&0&0.3059(1)&1.02(1)\\
O4&$4k$&1&0.25&0.7684(1)&0.7071(1)&0.79(1)\\
O5&$4k$&1&0.25&0.2495(1)&0.2792(1)&0.75(1)\\
O6&$4j$&1&0.0039(1)&0.5&0.2663(1)&0.82(1)\\
O7&$2f$&0.921(2)&0.25&0.5&0.5002(2)&0.43(2)\\
O8&$2e$&1&0.25&0&0.4996(12)&0.43(2)\\
\end{tabular}
\end{ruledtabular}
\end{table*}
%

\begin{table*}
\caption{\label{coordinate550pmmm}  The atomic occupancy, coordinates and the isotropic displacement parameters for $x$ = 0 sample obtained from Rietveld refinement on neutron powder diffraction data at 395 K. The refinement is performed using orthorhombic $Pmmm (a_p \times 2a_p \times 2a_p)$ structure.The lattice parameters and criteria of the refinement quality are: $a$ = 3.89832(1) \AA, $b$ = 7.90892(2)\AA, $c$ = 7.64649(2) \AA; $R_{wp}$ = 9.54\%. }
\begin{ruledtabular}
\begin{tabular}{ccccccc}
Atom&Site&Occupancy&$x$&$y$&$z$&$100U_{iso}(\AA^2)$\\ \hline
Ba&$2o$&1&0.5&0.2485(1))&0&0.63(2)\\
Pr&$2p$&1&0.5&0.2680(1)&0.5&1.45(3)\\
Co1&$2r$&1&0&0.5&0.2502(2)&0.94(4)\\
Co2&$2q$&1&0&0&0.2513(2)&0.75(3)\\
O1&$1a$&1&0&0&0&1.63(3)\\
O2&$1e$&1&0&0.5&0&0.95(3)\\
O3&$1g$&0.912(3)&0&0.5&0.5&0.94(4)\\
O4&$1c$&0.142(3)&0&0&0.5&0.94(4)\\
O5&$2s$&1&0.5&0&0.3041(1)&1.79(2)\\
O6&$2t$&1&0.5&0.5&0.2653(1)&1.46(2)\\
O7&$4u$&1&0&0.2373(1)&0.2876(1)&1.45(2)\\
\end{tabular}
\end{ruledtabular}
\end{table*}
%

\begin{table*}
\caption{\label{coordinate574}  The atomic occupancy, coordinates and the isotropic displacement parameters for $x$ = 0.24 sample obtained from Rietveld refinement on neutron powder diffraction data at 300 K. The refinement is performed using orthorhombic $Pmmm (a_p \times 2a_p \times 2a_p)$ structure.The lattice parameters and criteria of the refinement quality are: $a$ = 3.89832(1) \AA, $b$ = 7.90892(2) \AA, $c$ = 7.64649(2) \AA; $R_{wp}$ = 6.83\%. }
\begin{ruledtabular}
\begin{tabular}{ccccccc}
Atom&Site&Occupancy&$x$&$y$&$z$&$100U_{iso}(\AA^2)$\\ \hline
Ba&$2o$&1&0.5&0.2507(1)&0&0.09(2)\\
Pr&$2p$&1&0.5&0.2654(1)&0.5&0.56(3)\\
Co1&$2r$&1&0&0.5&0.2494(2)&0.49(3)\\
Co2&$2q$&1&0&0&0.2524(2)&0.41(4)\\
O1&$1a$&1&0&0&0&1.25(3)\\
O2&$1e$&1&0&0.5&0&0.84(3)\\
O3&$1g$&0.813(5)&0&0.5&0.5&0.45(4)\\
O4&$1c$&0.652(6)&0&0&0.5&1.11(4)\\
O5&$2s$&1&0.5&0&0.2874(2)&1.39(2)\\
O6&$2t$&1&0.5&0.5&0.2715(2)&1.18(2)\\
O7&$4u$&1&0&0.2484(1)&0.2763(1)&1.45(2)\\
\end{tabular}
\end{ruledtabular}
\end{table*}

The occurrence of TI-MIT in the PM phase makes it difficult to determine individual magnetic moments with neutron diffraction. Thus, we turned to lower temperatures and studied the magnetic structure in order to infer the magnetic moments in the PM phase from the nearby magnetic-ordered phase  [see Fig.~\ref{fig3}(c)]. Upon cooling the sample, a long-range magnetic ordering emerges starting at $T_{N1}$ = 255 K, which is identified by the reflection (111). This phase is henceforth referred to as Phase 1 and has the magnetic propagation vector $\bm{k}_{m1}$= (0, 0, 0). A new magnetic reflection ($11\frac{1}{2}$) appears starting at $T_{N2}$ = 237 K, which evidences a second magnetic-ordered phase. It is referred to as Phase 2 and has $\bm{k}_{m2}$  = (0, 0, $\frac{1}{2}$). Since the Pr$^{3+}$ spins become ordered below 20 K,\cite{Pr_ordering} the contribution to magnetic reflections  only  comes from only Co$^{3+}$ spins. The temperature dependence of these reflections is in the analogous pattern as those in RBaCo$_2$O$_{5.5}$ (R: Nd\cite{Fauth_2002}, Gd\cite{Chernenkov_2005}, Tb\cite{Plakhty_2005}, Dy\cite{Chernenkov_2007}, Ho\cite{Jorgensen_2008} and Y\cite{Khalyavin_2007}), where Phase 1 has been identified as a ferrimagnetic structure with nonzero net magnetic moment and Phase 2 as a antiferromagnetic structure with zero net magnetic moment. We adopted the ferromagnetic structure as the initial model in the Rietveld refinement and obtained the magnetic moments for the four sites as shown in Fig.~\ref{fig3}(d). As for the Phase 2, an antiferromagnetic structure for Phase 2 can be constructed from the ferrimagnetic magnetic structure in Phase 1 by reversing the magnetic moments in the neighboring Co layers linked by Pr layer along $c$ axis [see Fig.~\ref{fig3}(d)], which has been suggested by Plakhty $et$ $al.$\cite{Plakhty_2005} The relative difference in the magnitudes of moments points to the spin-state configuration of HS state for Co1(Pyr), IS state for Co2(Pyr) and Co4(Oct), and LS state for Co3(Oct).  Alternatively, the different moments can be attributed to the spin-state disproportionation; namely, each Co site resides in a HS/LS mixed state and the ratio of HS/LS differs among the four sites.

The pronounced difference between the Shannon effective ionic radii\cite{Shannon_1976} of the LS  Co$_{3+}$ ion (0.545 \AA) and the HS one (0.61 \AA) in octahedron coordination also helps to identify the spin states. Fig.~\ref{fig3}(e) shows that the average Co3(Oct)$-$O and Co4(Oct)$-$O bond lengths below $T_{MI}$ amount to a greater than 0.06 Å difference, consistently corroborating the conclusion that Co4(Oct) resides in HS/IS spin state whereas Co3(Oct) resides in LS state. Therefore, a long-range SSO is identified in the $Pmma$ phase of PrBaCo$_2$O$_{5.5}$, which has also been found in other rare-earth families, RrBaCo$_2$O$_{5.5}$  (R: Nd\cite{Fauth_2002}, Gd\cite{Chernenkov_2005}, Tb\cite{Plakhty_2005}, Dy\cite{Chernenkov_2007}, Ho\cite{Jorgensen_2008} and Y\cite{Khalyavin_2007}). The spin-state configuration in the high-temperature $Pmmm$ phase cannot be directly determined as there is no magnetic-ordered phase nearby in phase space. Alternatively, the analysis presented below on the magnetic structure of the hole-doping sample will shed light on the spin states in this high-temperature phase. We can now understand that the crystal structure change from $Pmma$ to $Pmmm$ corresponds to the melting of SSO.

\begin{figure*}
\includegraphics[width=0.8\textwidth]{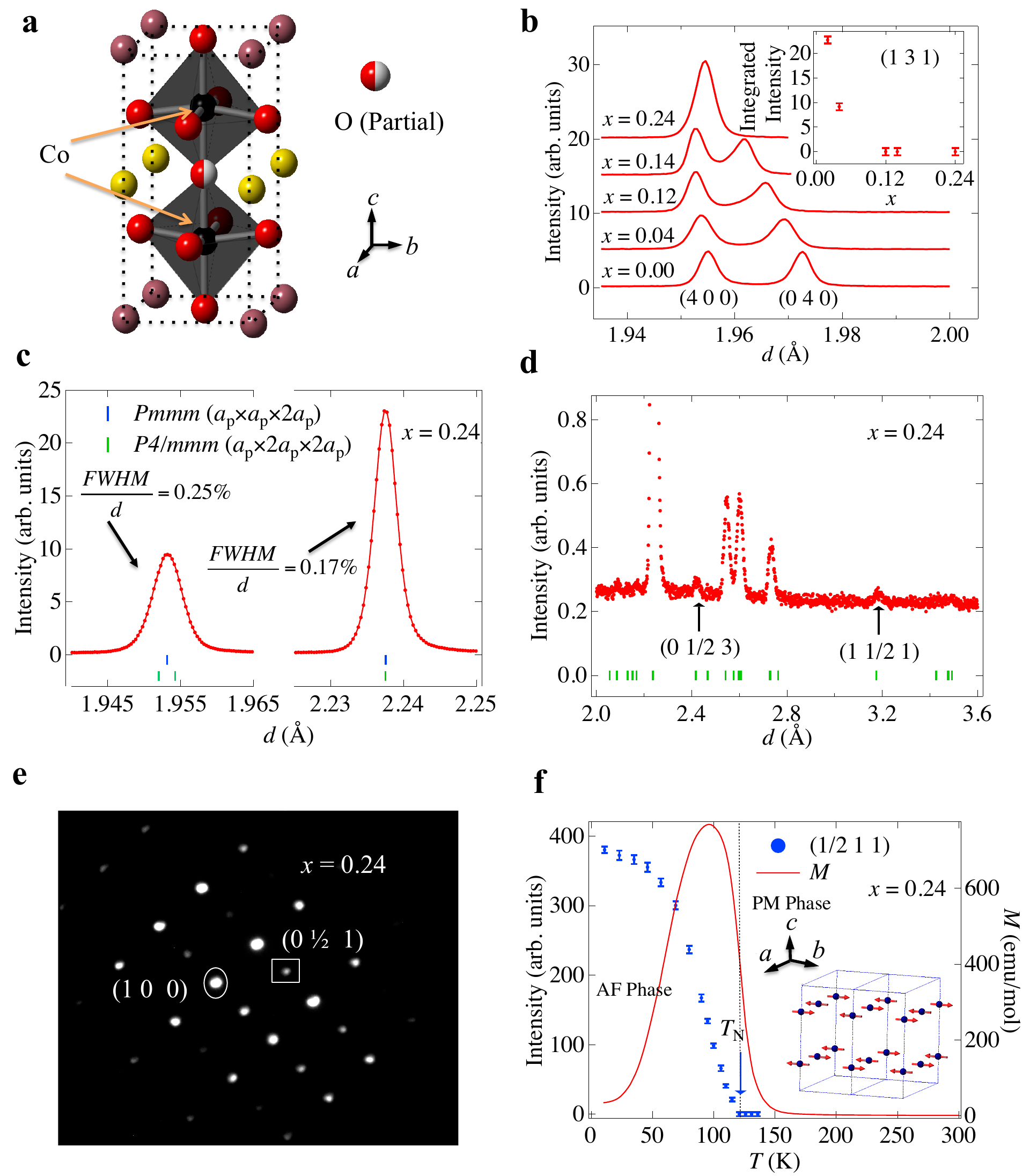}
\caption{\label{fig4}  Crystal structure model (a), results of pulsed neutron powder diffraction (b, c, d, f), electron diffraction (e) and inverse magnetic susceptibility for PrBaCo$_2$O$_{5.5+x}$ (HDI-MIT).  (a)  The $P4/mmm (a_p \times a_p \times 2a_p)$ unit cell with the random distribution of oxygen vacancies in the  PrO$_x$ layer, leading to only one nonequivalent crystallographic Co site. (b)  The diffraction pattern of the reflections 400 and 040, as well as the integrated intensity of reflection 131 (inset),  which are indexed by the unit cell $(2a_p \times 2a_p \times 2a_p)$,  as a function of hole-doping fractions $x$ at 300 K.  The results suggest that crystal structure transforms from the orthorhombic $Pmma (2a_p \times 2a_p \times 2a_p)$ to the orthorhombic $Pmmm (a_p \times 2a_p \times 2a_p)$ at $x$ =0.12. (c)  Comparison of the 200 reflection (left) with the 112 reflection (right) indexed by the tetragonal $(a_p \times a_p \times 2a_p)$ unit cell. The 200 reflection splits into two reflections, whereas the 112 reflection remains to be single in the framework of the orthorhombic $(a_p \times 2a_p \times 2a_p)$ unit cell. (d, e)  The superlattice reflections with respect to the $(a_p \times a_p \times 2a_p)$ unit cell observed in both neutron and electron diffraction patterns of $x$ = 0.24 sample at 300 K. (f)  Magnetization (right axis) and integrated intensity (left axis) for the magnetic reflection $\frac{1}{2} 1 1$ indexed by the unit cell $(a_p \times 2a_p \times 2a_p)$ as a function of temperature for $x$ = 0.24, marking the antiferromagnetic phase (AF phase) with $T_{N}$ = 120 K.}
\end{figure*}

\subsection{Crystal and magnetic structures of PrBaCo$_2$O$_{5.5+x}$ ($x > 0$)}
We now discuss the HDI-MIT in the $0 < x \le 0.24$ samples, which covers the transformation at $x$ = 0.12 from the insulting phase to the poor-metallic phase [see inset of Fig.~\ref{fig1_1}(b)]. The insulating side of the HDI-MIT starts at the stoichiometric end member PrBaCo$_2$O$_{5.5}$ with the $Pmma (2a_p \times 2a_p \times 2a_p)$ structure. Hole doping introduces oxygen ions into the empty site (pyramid) of in RO$_x$ layer as shown in  Fig.~\ref{fig1}(b). The inset of Fig.~\ref{fig4}(b) shows the integrated intensity of the (131) neutron powder diffraction peak at 300 K, which declines as a function of hole-doping fraction $x$, reflecting a decreasing contribution from the $Pmma (2a_p \times 2a_p \times 2a_p)$ structure. The intensity finally vanishes at x = 0.12, which indicates a structural phase transition. Previous studies\cite{Taskin_2005, Frontera_2004} asserted that for $x > 0.1$, RBaCo$_2$O$_{5.5+x}$ crystallizes into the tetragonal $P4/mmm (a_p \times a_p \times 2a_p)$ structure with only one nonequivalent crystallographic site for Co ion, corresponding to the partial and random distribution of oxygen vacancies in the RO$_x$ layer [see Fig.~ \ref{fig4}(a)]. However, Fig.~\ref{fig4}(b) shows that we see the 040 and 400 reflections in, whose positions are directly related to the lattice constants $a$ and $b$, remain to be separate until $x$ = 0.14, unambiguously suggesting that the structure transforms into orthorhombic $Pmmm (a_p \times 2a_p \times 2a_p)$, which corresponds to the disproportionate distribution of oxygen vacancies in PrO$_x$ layer [$i.e.$, oxygen ions prefer the right site to left one in  Fig.~\ref{fig1}(a)]. Although the 040 and 400 reflections seem to merge into a single peak in Figure 3b at $x$ = 0.24, the ratio of full width at half maximum (FWHM) to the $d$-spacing is 0.25\%, which is broader than the other single peak [$\frac{FWHM}{d}$ $\approx 0.17\%$; see in Fig.~ \ref{fig4}(c)], indicating an orthorhombic distortion. Additional evidences for the $Pmmm (a_p \times 2a_p \times 2a_p)$ structure at $x$ = 0.24 is provided by both neutron and electron diffraction, in which superlattice reflections along $b$ axis with respect to the $(a_p \times a_p \times 2a_p)$ unit cell are observed [see Fig.~ \ref{fig4}(d) and Fig.~ \ref{fig4}(e), respectively]. The results from Rietveld refinement is shown in Table~\ref {coordinate574}. 
Therefore, we conclude that the crystal structure phase transition across the HDI-MIT is from $Pmmm (a_p \times 2a_p \times 2a_p)$ to $Pmma (2a_p \times 2a_p \times 2a_p)$, which is identical to that across the TI-MIT.

\begin{table*}
\caption{\label{tab:BV} The basis vectors (BVs) of irreducible representations (IRs) for the Co1 site at 2$r$: [Atom I-(0, $\frac{1}{2}$, $z$); Atom II-0, $\frac{1}{2}$, -$z$)] of  $Pmmm (a_p \times 2a_p \times 2a_p)$ crystal structure with propagation vector  $\bm{k}_{m}$ = ($\frac{1}{2}$, 0, 0). The BVs for Co2 site at 2$q$: [Atom I-(0, 0, $z$); Atom II-(0, 0, -$z$)] are the same as those for Co1 site.}
\begin{ruledtabular}
\begin{tabular}{cccccccc}
 &&\multicolumn{3}{c}{\textrm{Atom I}}&\multicolumn{3}{c}{\textrm{Atom II}}\\
IR&BV&$m_x$&$m_y$&$m_z$&$m_x$&$m_y$&$m_z$\\ \hline
$\Gamma_2$&$\phi_1$&0&0&4&0&0&-4\\
$\Gamma_3$&$\phi_2$&4&0&0&4&0&0\\
$\Gamma_4$&$\phi_3$&0&4&0&0&-4&0\\
$\Gamma_5$&$\phi_4$&0&4&0&0&4&0\\
$\Gamma_6$&$\phi_5$&4&0&0&-4&0&0\\
$\Gamma_7$&$\phi_6$&0&0&4&0&0&4\\
\end{tabular}
\end{ruledtabular}
\end{table*}
%

\begin{table*}
\caption{\label{tab:moments} Refined magnetic moments ($\mu_B$) for the two crystallographic cobalt sites of $x$ = 0.24 sample based on the basis vectors from Table 1 in the main text. $R_M$ is the magnetic R-factor for Rietveld refinement. The resulting magnetic structure models are shown in Fig.~\ref{fig:model}.}
\begin{ruledtabular}
\begin{tabular}{ccccccc}
&$\Gamma_2$&$\Gamma_3$&$\Gamma_4$&$\Gamma_5$&$\Gamma_6$&$\Gamma_7$\\ \hline
$M_x$-Co1&0&0.10(1)&0&0&1.27(1)&0\\
$M_y$-Co1&0&0&1.31(1)&0.10(1)&0&0\\
$M_z$-Co1&1.22(1)&0&0&0&0&0.44(3)\\
$M_x$-Co2&0&-0.10(1)&0&0&-1.27(1)&0\\
$M_y$-Co2&0&0&-1.31(1)&-0.10(1)&0&0\\
$M_z$-Co2&-1.22(1)&0&0&0&0&-0.44(3)\\
$R_M$(\%)&21.7&189.0&8.6&189.0&9.1&112.0\\
\end{tabular}
\end{ruledtabular}
\end{table*}

We also analyzed the magnetic structure of $x$ = 0.24 sample to infer the spin states of Co$^{3+}$ ions at higher temperatures because the HDI-MIT also occurs in the PM phase [see Fig.~ \ref{fig4}(f)]. Our synchrotron XRD experiment reveals that the $Pmmm$ structure is preserved at temperatures down to about 30 K. As shown in Fig.~\ref{fig4}(f), the magnetic ordering phase with propagation vector $\bm{k}_{m}$ = ( $\frac{1}{2}$, 0, 0) can be identified in neutron diffraction pattern by the appearance of the ( $\frac{1}{2}$11) magnetic reflection at $T_{N}$ = 120 K, whose intensity gradually increases with decreasing temperature. According to Landau theory, the magnetic ordering from such a second-order phase transition will involve a single irreducible representation (IR). Therefore, given the $Pmmm$ crystal structure with $\bm{k}_{m}$ = ($\frac{1}{2}$, 0, 0), representation analysis allows us to determine the possible IRs and the corresponding basis vectors (BVs) for magnetic structure. The Co ions are located in two different crystallographic positions, i.e., Co1 at 2$r$: [I$-$(0, 1/2, $z$); II$-$(0, 1/2, -$z$)] and Co2 at 2$q$: [I$-$(0, 0, $z$); II$-$(0, 0, -$z$)]. The decomposition of magnetic representation using the program $SARAh$\cite{Wills_2000} gives the same IRs for Co1 and Co2 site, $i.e.$,
\begin{eqnarray}
\Gamma_{Mag}=&&0\Gamma_1^{(1)}+1\Gamma_2^{(1)}+1\Gamma_3^{(1)}+1\Gamma_4^{(1)}+1\Gamma_5^{(1)}+1\Gamma_6^{(1)}\nonumber\\
&&+1\Gamma_7^{(1)}+0\Gamma_8^{(1)}
\end{eqnarray}

\begin{figure}
\includegraphics[width=\columnwidth]{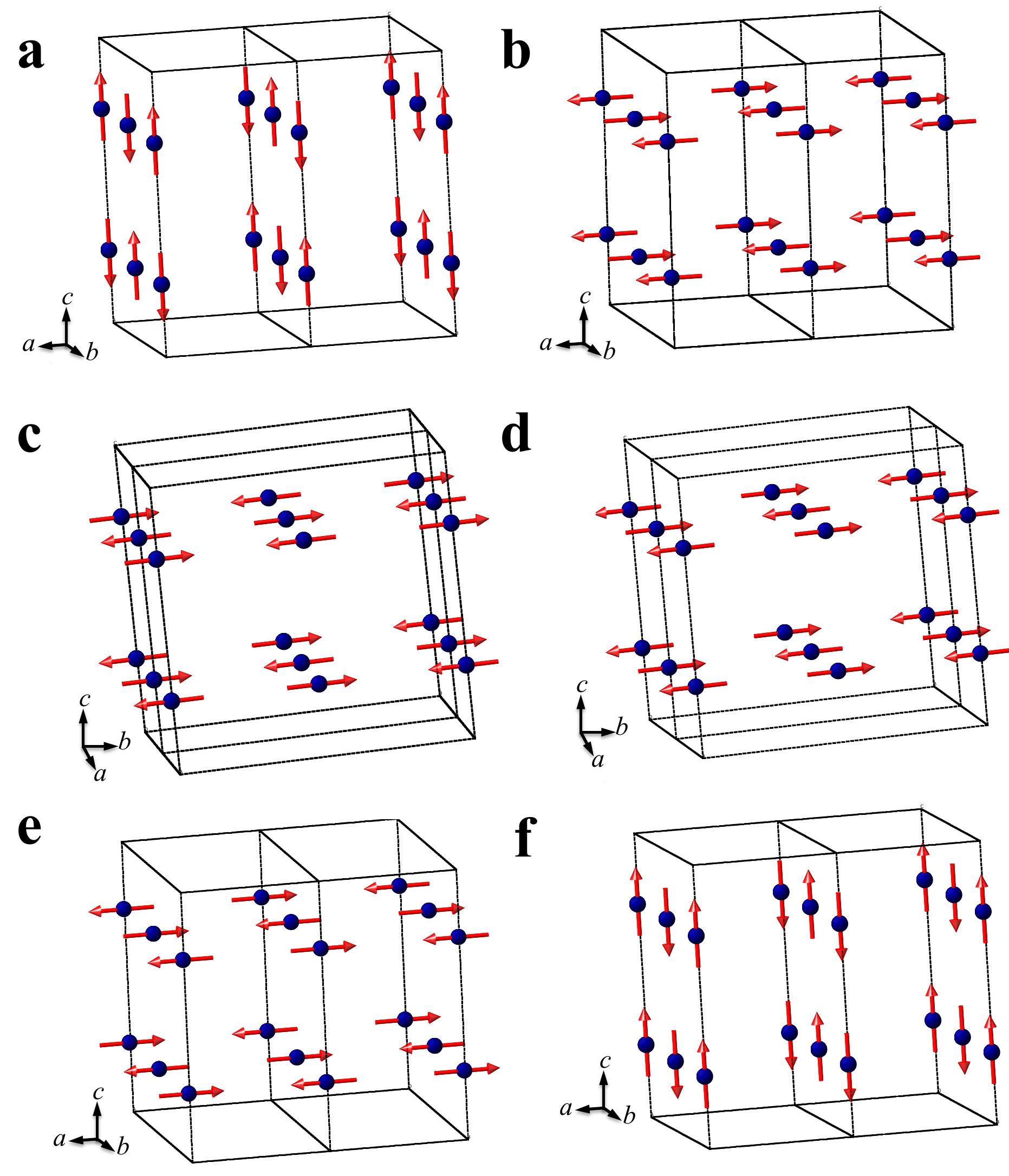}
\caption{\label{fig:model}  Schematic drawing of the refined magnetic structures from Table~\ref{tab:moments}. (a)  $\Gamma_2$. (b)  $\Gamma_3$. (c)  $\Gamma_4$. (d)  $\Gamma_5$. (e)  $\Gamma_6$. (f)  $\Gamma_7$. The magnetic unit cell is double of crystal unit cell along the $b$ axis.}
\end{figure}

\begin{figure}
\includegraphics[width=0.8\columnwidth]{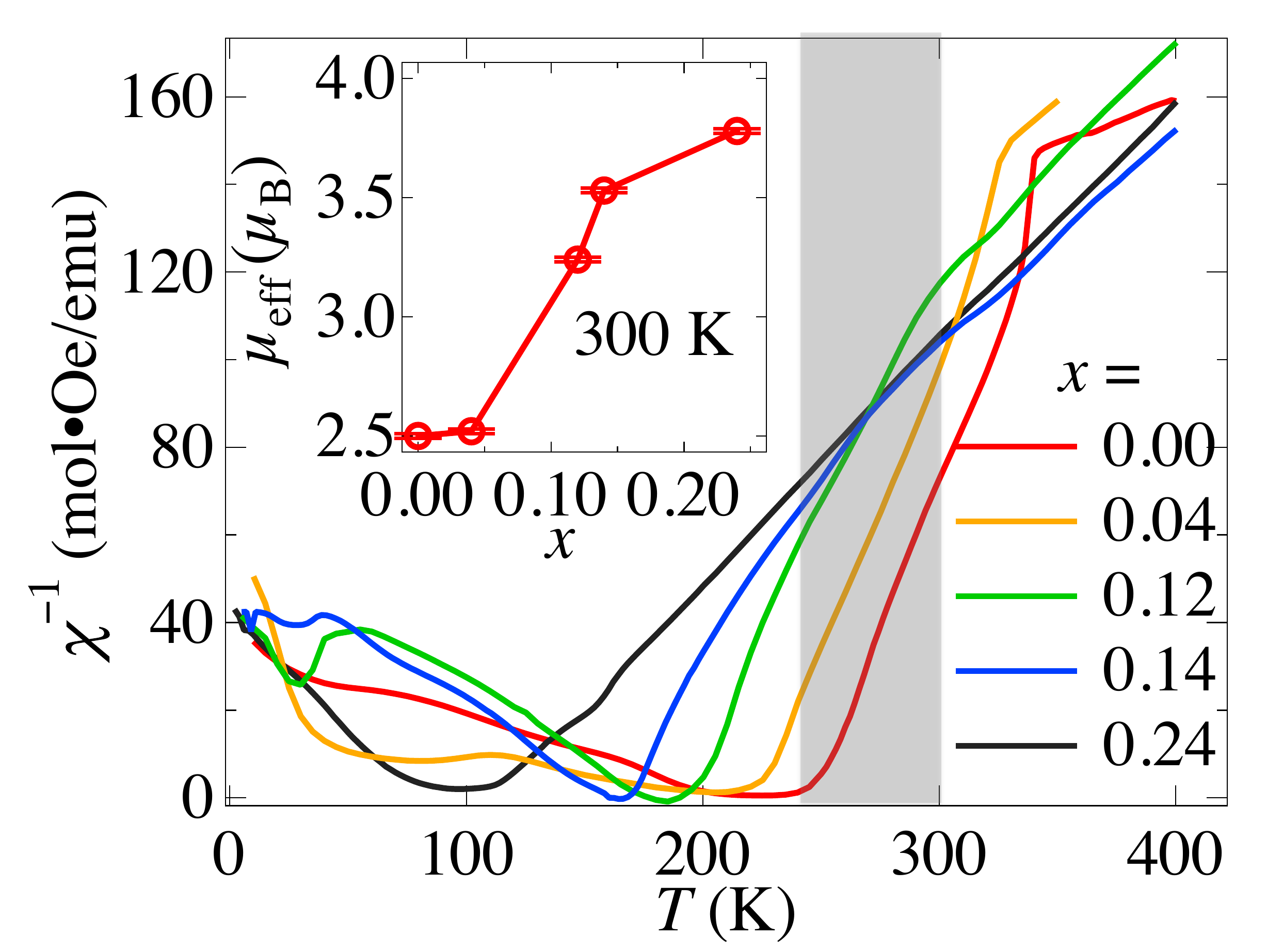}
\caption{\label{fig5}  The inverse magnetic susceptibility $\chi^{-1}$ as a function of temperature for various hole-doping fractions $x$. The slope increases in paramagnetic phase (shaded region) with increasing $x$, corresponding to the x-dependence of effective magnetic moments  $\mu_{eff}$ at 300 K as obtained from a Curie-Weiss fit within the shaded region (see inset).}
\end{figure}

Also, the same BVs for both sits are obtained by means of projection operator as given in Table~\ref{tab:BV} and the corresponding magnetic structures are shown in Fig.~\ref{fig:model}. 
Through Rietveld refinement (see Table~\ref{tab:moments}), we find an G-type antiferromagnetic structure with the Co-ion spins pointing along either $a$ axis ($\Gamma6$) or $b$ axis ($\Gamma4$) as shown in Figure 3f, and the magnitude of magnetic moment $M$ at 11 K is determined to be about 1.3 $\mu_B$. The x-ray absorption spectroscopy studies by Medling $et$ $al.$\cite{Medling_2012, Medling_2013} indicates that in La$_{1-x}$Sr$_x$CoO$_3$ the doping induced holes are nearly equally distributed over the Co and O atoms, resulting in coexistence of the Co$^{4+}$ ions and the magnetic O$^-$ ions. Accordingly, we expect the $x = 0.24$ sample contains statistically 12\% Co$^{4+}$ ions and 12\% magnetic O$^-$ ions. Since 12\% of Co4+ ions ($S = \frac{1}{2}$, $t_{2g}^5e_g^0$) and 12\% of O$^-$ ions ($S = \frac{1}{2}$) contribute at most $\mu_B$ to $M$, the major component must come from the Co$^{3+}$ ions; namely, the Co$^{3+}$ ions in the two Co sites of $Pmmm (a_p \times 2a_p \times 2a_p)$ structure are all in IS ($S = 1$, $t_{2g}^4e_g^2$) states. Comparing with the spin-state configuration in the $Pmma (2a_p \times 2a_p \times 2a_p)$ structure at $x = 0$, we can see that by increasing the hole-doping fraction $x$, the LS and HS Co$^{3+}$ ions undergo the SST into IS state so that the SSO in $Pmma$ model is melted. Further evidence supporting this conclusion is a jump of the Curie-Weiss effective magnetic moment $\mu_{eff}$ from 2.52(1) $\mu_B$ ($x = 0.04$) to 3.24(1) $\mu_B$ ($x = 0.12$), as shown in Fig.~\ref{fig5}. 
Because only 8\% increase in the total number of Co$^{4+}$ and O$^-$ ions cannot account for the 0.72 $\mu_B$ difference, some of the Co3+ ions must convert from the LS to IS state. As a result, the hole doing to PrBaCo$_2$O$_{5.5}$ induced the SST, resulting in spin-state order-disorder transition, which is in the same manner as increasing the temperature in PrBaCo$_2$O$_{5.5}$.

\subsection{Soft x-ray absorption spectroscopy (XAS) on both MITs}
\begin{figure}
\includegraphics[width=0.8\columnwidth]{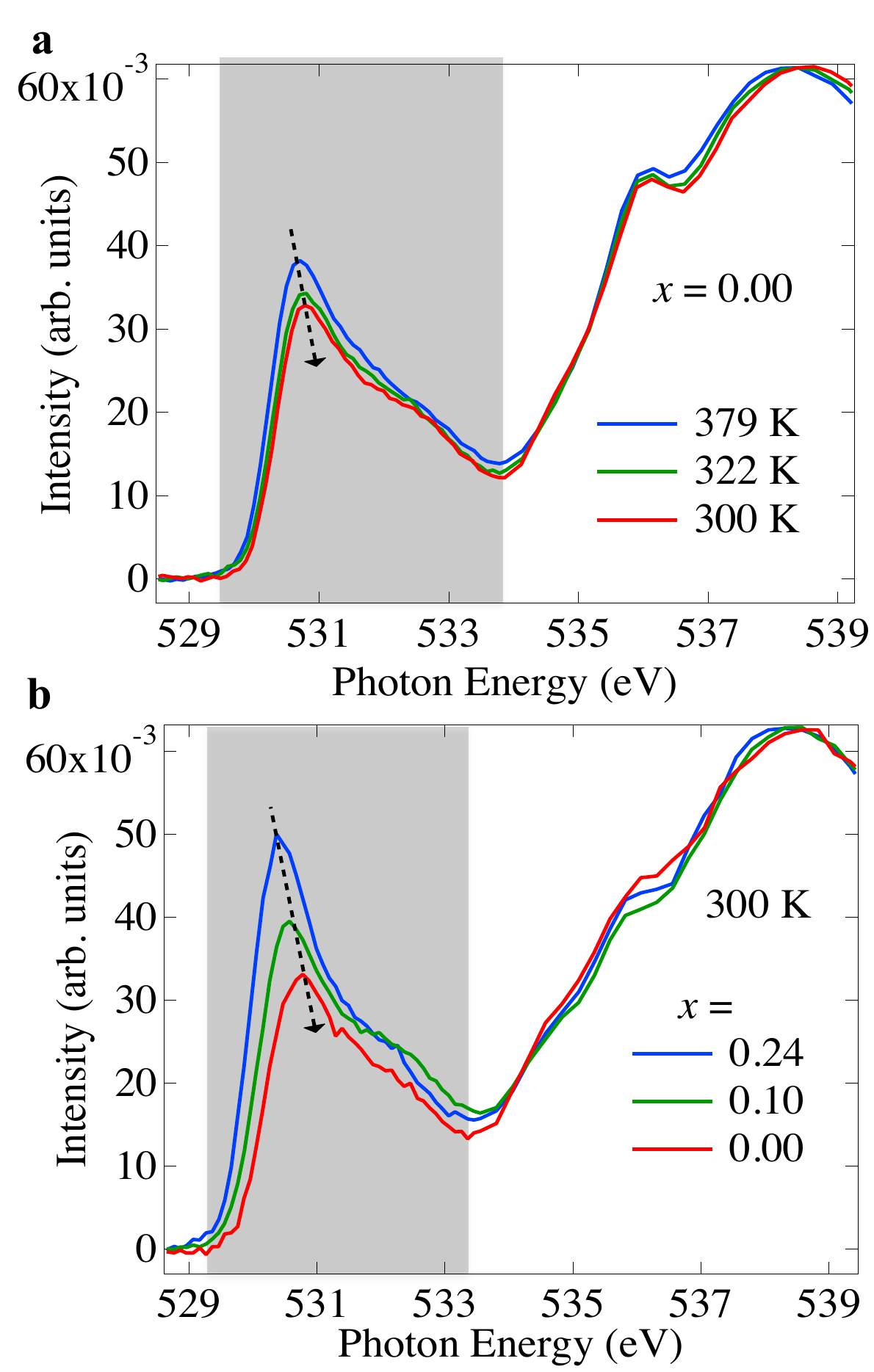}
\caption{\label{fig6}  Soft x-ray absorption spectra (XAS) at oxygen $K$-edge for PrBaCo$_2$O$_{5.5}$ (a, TI-MIT) and for PrBaCo$_2$O$_{5.5+x}$ (b, HDI-MIT).  The intensities are normalized at about energy 539 eV after subtracting a small constant background.  Upon decreasing either the temperature or hole-doping fraction, the threshold of the spectra shifts to higher energy side as shown by the arrow in the shaded region, indicating the common nature for both TI- and HDI-MITs}
\end{figure}

The common behavior of the two MITs also appears in the spectra of oxygen K-edge soft x-ray absorption spectroscopy (XAS), see Fig.~ \ref{fig6}(a) and (b). The shaded region from 529 to ~534 eV is due to the transition from the O $1s$ core level to the O $2p$ orbitals that are hybridized with the unoccupied Co $3d$ $t_{2g}$ and $e_g$ states, reflecting the density of states of the bottom of the conduction band. The broad structures above 534 eV are due to Ba $3d$, Co $4s$, and Pr $4f$ related bands.\cite{Hu_2004} Upon decreasing either the temperature or hole-doping fraction, the threshold of spectra shifts in to the higher energy side as shown by the arrows in Figure 4a and Figure 4b, which indicates that the band gap widens for both MITs. The hole-doping dependence of spectra is analogous to that of La$_{1-x}$Sr$_x$CoO$_3$, where a part of the holes were found to reside on the O atoms, reflecting a strong Co$-$O hybridization.\cite{Medling_2012, Medling_2013} Therefore, the shift of absorption edge in PrBaCo$_2$O$_{5.5+x}$ indicates that either temperature or hole doping causes the variation in the Co$-$O hybridization. Consequently a change in the strength of crystal field on Co$^{3+}$ ions occurs, which serves as the driven force for spin-state transition. 

\section{Discussion}
The first discovery of present study is that hole doping to the stoichiometric compound PrBaCo$_2$O$_{5.5}$ induces melting of SSO, coinciding with the insulator$-$metal transition. The melting of SSO triggers the crystal structure change from $Pmma (2a_p \times 2a_p \times 2a_p)$ to $Pmmm (a_p \times 2a_p \times 2a_p)$, which is identical to what occurs in the TI-MIT. As $Pmma (2a_p \times 2a_p \times 2a_p)$ belongs to the maximal subgroup of $Pmmm (a_p \times 2a_p \times 2a_p)$, the symmetry breaking near the phase-transition point (at $T_{MI}$ or $x_{MI}$) suggests that the electronic structure changes. Such electronic change manifests itself in somewhat identical manner regardless of whether it is driven by tuning temperature or hole doping: a rapid drop in resistivity in PM phase, a rise of the effective magnetic moments, and a shift in the soft x-ray absorption edges.

\begin{figure*}
\includegraphics[width=0.8\textwidth]{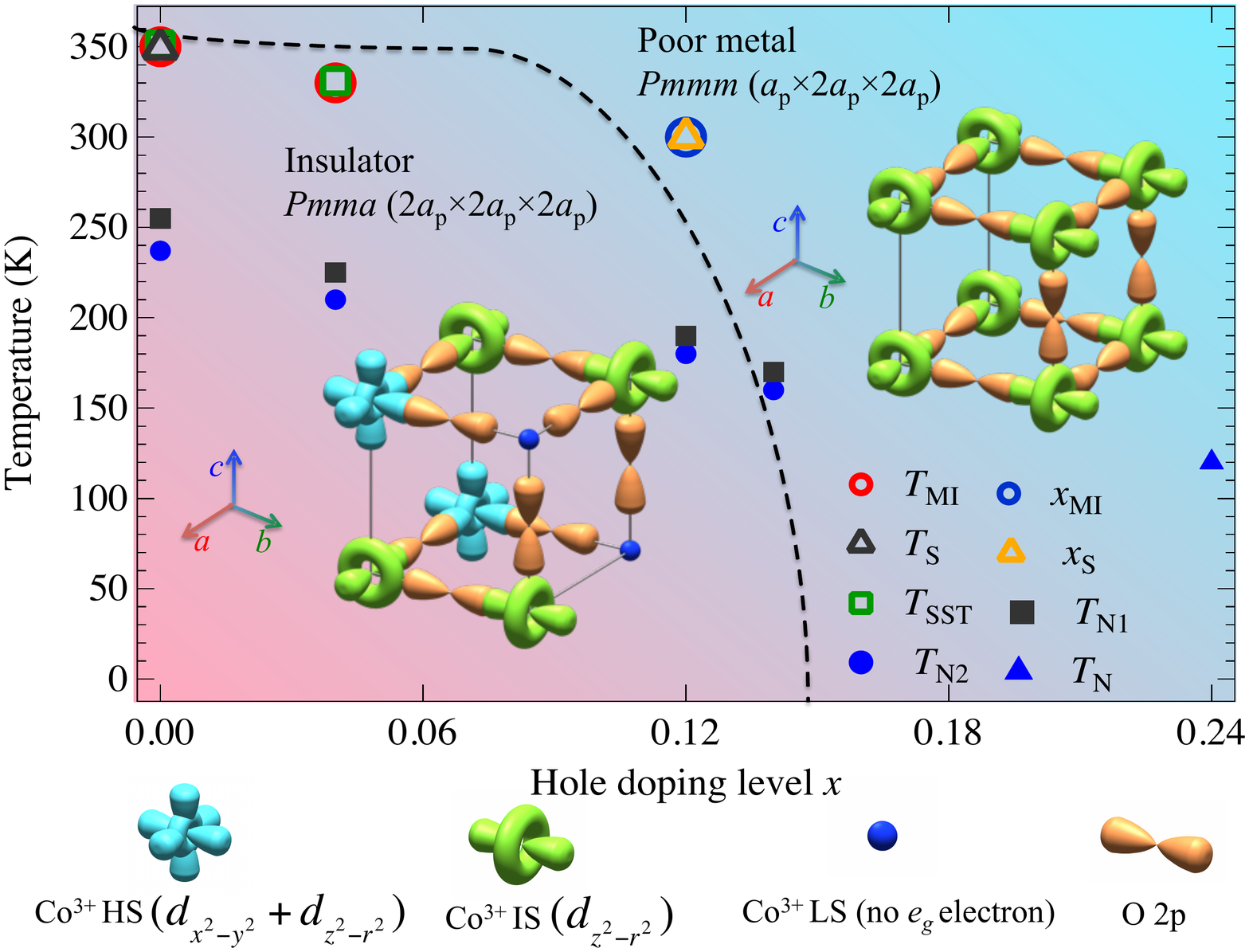}
\caption{\label{fig7}  Phase diagram of PrBaCo$_2$O$_{5.5+x}$ featuring the common nature for both the temperature- and hole-doping-induced MITs. The MITs are parameterized by $T_{MI}$ and $x_{MI}$, which are determined from the resistivity measurements (see Fig.~\ref{fig1_1}(b). The crystal structure transition temperatures $T_{S}$ and hole doping fractions $x_{S}$ are determined by various diffraction techniques probes (see Fig.~\ref{fig3}(e) and ~\ref{fig4}(b), respectively). $T_{SST}$ corresponds to the SST, which is identified by the change in the slope of the inverse susceptibility vs temperature [see Fig.~\ref{fig5}]. $T_{N1}$, $T_{N2}$ and $T_{N}$, which are determined from the magnetization measurements [see Fig.~\ref{fig5}] and neutron diffraction measurements [see Fig.~\ref{fig3}(c) and Fig.~\ref{fig4}(f)], denote the magnetic ordering transition temperature in the insulating phase and the poor metallic phase, respectively. The schematic electronic structures are arranged in the form of a pseudocubic perovskite subcell with a Pr ion at body center. It represents the entire electronic structure because this subcell shares the CoO layer with that of the Ba-ion body center. The spin states of Co$^{3+}$ ions are denoted by the occupancy of $e_g$ orbital, which clearly illustrates the SSO along the $a$ axis in insulating phase. The pure spin state at each Co site may also be interpreted by the HS/LS mixed state with different ratios of HS/LS among the Co sites. The orbital orientations are inferred by comparing the individual Co$-$O bond lengths along the $a$, $b$, $c$ axes. Thus, these orientations involve some uncertainty.}
\end{figure*}

The combined results are depicted in Fig.~\ref{fig7}, 
which shows a schematic phase diagram at high-temperature regions (i.e., PM phase) for the PrBaCo$_2$O$_{5.5+x}$ system. Although the data are insufficient to fully characterize the exact phase boundary, the overall picture is consistent with all the experimental data from studies of structural refinements, magnetic and transport properties, and electronic band structure. The results demonstrate the uniform IS distribution (or HS/LS mixture) in the metallic $Pmmm (a_p \times 2a_p \times 2a_p)$ phase transforms to the SSO distribution in the insulating $Pmma (2a_p \times 2a_p \times 2a_p)$ phase. The SSO configuration consists of HS state for Co1(Pyr), IS state for Co2(Pyr) and Co4(Oct), and LS state for Co3(Oct); alternatively, it is composed of HS/LS mixed states with different ratios of HS/LS among the four sites. The SSDF stems from the competition between the crystal field and Hund coupling18, which in case of cobaltite are associated with the hybridization of the Co $3d$ with the O $2p$ orbitals. So the symmetry breaking of lattice leads to different local Co$-$O hybridizations, which results in the SSO in PrBaCo$_2$O$_{5.5+x}$. Here, we argue that the Co$-$O hybridization is the primary factor responsible for the MITs, whereas the magnetic Co$-$Co direct exchange and Co$-$O$-$Co super exchange interactions are secondary. This explains why the TI-MIT occurs above the magnetic ordering temperature and why SSO is preserved even in the PM phase. The corroborative evidence for the strong Co$-$O hybridization is also found in the high epitaxial thin films of LaBaCo$_2$O$_{5.5+x}$, where the interface strain significantly alters the electronic transport properties.\cite{Ma_2014, Ma_2013, Liu_2012}

Recently, some first-principle calculations\cite{Iron_spin_state_2013, Iron_spin_state_prl_2013, Iron_spin_state_2011} indicate that the Fe$-L$ ($L$: ligand) hybridization in iron-based superconductors induces a superposition of different spin states of Fe$^{2+}$ ion. The weight of each spin sate is modulated by this Fe$-L$ hybridization, which can account for the unusual temperature dependence of the PM susceptibility.\cite{Imai_2009, Iron_SST_2013, Iron_SST_2012} Comparatively, our results demonstrate a different form of spatial spin-state distribution ($i.e.$, SSO) through the complex Co$-$O hybridization, which also leads to the unusual temperature-dependence of PM susceptibility. Therefore, we expect the SSO may occur in other systems with SSDFs, including the iron-based superconductors, and it can correlate with many novel magnetic and transport properties.

\section{Conclusion}
In conclusion, we present convincing evidences from multi-probes (neutron, electron and x-rays) that hole doping to PrBaCo$_2$O$_{5.5}$ melts the spin-state ordering (SSO) of Co$^{3+}$ ions in conjunction with an insulator-metal transition, which is in the same manner as the temperature-induced metal-insulator transition (MIT).  A unified mechanism is proposed to dominate the temperature- and hole-doping-induced MITs in the PrBaCo$_2$O$_{5.5+x}$ system, $i.e.$, the symmetry breaking coupled with the SSO. 

\begin{acknowledgments}
The neutron measurements were finaced by proposal No. 2014S05 of the S-type project of KEK and XAS experiment under proposal No. 2014V001. We thank Prof. Y. Noda and Dr. J. Zhang for helpful discussion. We also appreciate M. Ishikado, T. Moyoshi, M. Shioya and T. Muroya for their assistance with the experiments.
\end{acknowledgments}

\bibliography{Reference}

\end{document}